\documentclass[prc,twocolumn,showpacs,preprintnumbers,amsmath,amssymb]{revtex4}
\usepackage{epsfig}
\usepackage{graphicx}
\usepackage{bm}
\usepackage[dvips]{color}
\usepackage{colordvi}
\def\ci{\cite}

\def\lsim{\buildrel < \over {_{\sim}}}

\newcommand{\beq}{\begin{equation}}
\newcommand{\eeq}{\end{equation}}
\newcommand{\be}{\begin{eqnarray}}
\newcommand{\ee}{\end{eqnarray}}

\begin{document}
\title{Microscopic calculations of transport properties of 
neutron matter}
\author{Omar Benhar$^{1,2}$}
\author{Artur Polls$^{3}$}
\author{Marco Valli$^{1,2}$}
\author{Isaac Vida\~na$^{4}$}
\affiliation
{
$^1$ INFN, Sezione di Roma. I-00185 Roma, Italy\\
$^2$ Dipartimento di Fisica, Universit\`a ``La Sapienza''. I-00185 Roma, Italy \\
$^3$ Departament d'Estructura i Constituents de la Mat\`eria. E-08028 Barcelona, Spain \\
$^4$ Centro de F\'isica Computacional, Department of Physics, University of Coimbra, 3004-516 Coimbra, Portugal \\
}
\date{\today}
\begin{abstract}
We discuss the results of calculations of the shear viscosity and thermal 
conductivity of pure neutron matter, carried out within the Landau-Abrikosov-Khalatnikov formalism.
The probability of neutron-neutron collisions in the nuclear medium has been 
obtained from a realistic potential, using both the correlated basis function  
and the $G$-matrix approach. The results of our work indicate that medium 
modifications of nucleon-nucleon scattering are large, their inclusion 
leading to a dramatic enhancement of the transport coefficients.
On the other hand, the results obtained from the two theoretical schemes appear to 
be in fairly good agreement.
\end{abstract}
\maketitle

\section {Introduction}
\label{intro}

The knowledge of transport properties of neutron matter is relevant to 
the understanding of a variety of neutron star properties.
Viscosity plays a crucial role in determining the onset of the gravitational-wave driven
instability, associated with the excitation of $r$-modes, in rapidly rotating
stars \cite{AndKok}, while thermal conductivity is one of the factors determining 
neutron star cooling \cite{cooling}.

Unlike the equation of state (EOS), which is generally obtained
from realistic dynamical models, strongly constrained from nuclear systematics and
nucleon-nucleon scattering data, the non-equilibrium properties of neutron star matter 
are often studied using oversimplified models of the nucleon-nucleon (NN) interaction.

The main difficulty involved in the calculation of the transport coefficients within 
the formalism originally developed by Abrikosov and Khalatnikov \cite{ak}, 
based on Landau theory of normal Fermi liquids \cite{baym-pethick}, is the determination of
the nucleon-nucleon (NN) collision probability in the nuclear medium. 
Most studies of the transport properties of neutron star matter have circumvented this
 problem neglecting medium modifications of the NN cross sections altogether, and using 
the measured NN scattering phase shifts to construct the collision 
probability \cite{flowi1,flowi2,haensel}.

Nuclear many body theory provides a consistent framework to obtain the in-medium 
NN cross section and the transport coefficients of nuclear matter from realistic NN potentials, 
using either the $G$-matrix \cite{wambach} or the correlated basis function (CBF) \cite{bv}
formalism. In both approaches one can define a well behaved effective
interaction, suitable for use in standard perturbation theory in the Fermi gas basis
and allowing for a {\em consistent} treatment of equilibrium and non-equilibrium properties 
\cite{wambach,shannon,bv}. 

In this paper we discuss the results of calculations of the 
 shear viscosity and thermal conductivity of pure neutron matter, carried out 
using the CBF and $G$-matrix effective interactions.  

In Section \ref{veff_theory}, after outlining the elements of nuclear many body theory, 
we analyze the main features of the CBF and $G$-matrix effective interactions, while Section \ref{NNscatt}
is devoted to the discussion of the in-medium NN cross section in the kinematical 
setup relevant to the calculation of the transport coefficients. The main 
features of the Abrikosov-Khalatnikov formalism are reviewed in Section \ref{transport}, where 
we also present the results of numerical calculations. Finally, in Section \ref{conclusions} we summarize our
findings and state the conclusions.
\section{Effective interactions in Nuclear Many Body Theory}
\label{veff_theory}

Nuclear many body theory (NMBT) is based on the tenet that nuclei can be
described in terms of point like nucleons, whose dynamics are dictated by 
the hamiltonian 
\beq
H = \sum_i \frac{{\bf k}^2_i}{2m} + \sum_{j>i} v_{ij} + \sum_{k>j>i} V_{ijk}  \ ,
\label{hamiltonian}
\eeq
${\bf k}_i$ and $m$ being the momentum of the $i$-th nucleon and its mass, respectively. 

The nucleon-nucleon (NN) potential $v_{ij}$ 
reduces to the Yukawa one-pion exchange potential at large
distances, while its behavior at short and intermediate range is determined by
a fit of deuteron properties and NN scattering phase shifts.
The state-of-the-art NN parametrization referred to as Argonne $v_{18}$ potential \ci{av18}
is written in the form
\beq
v_{ij}=\sum_{n=1}^{18} v_{n}(r_{ij}) O^{n}_{ij} \ .
\label{av18:1}
\eeq
In the above equation
\beq
O^{n \leq 6}_{ij} = [1, (\bm{\sigma}_{i}\cdot\bm{\sigma}_{j}), S_{ij}]
\otimes[1,(\bm{\tau}_{i}\cdot\bm{\tau}_{j})]
\label{av18:2}
\eeq
where $\bm{\sigma}_{i}$ and $\bm{\tau}_{i}$ are Pauli matrices acting in
spin and isospin space, respectively, and
\beq
S_{ij}=\frac{3}{r_{ij}^2}
(\bm{\sigma}_{i}\cdot{\bf r}_{ij}) (\bm{\sigma}_{j}\cdot{\bf r}_{ij})
 - (\bm{\sigma}_{i}\cdot\bm{\sigma}_{j}) \ .
\label{def:S12}
\eeq
The operators corresponding to $n=7,\ldots,14$ are associated with the
non-static components of the NN interaction, while those
 corresponding  to $n=15,\ldots,18$ account for small charge symmetry violations.
Being fit to the full Nijmegen phase shifts data base, as well as to
low energy scattering parameters and deuteron properties, the Argonne $v_{18}$ potential
provides an accurate description of the scattering data by construction.

The three-nucleon potential $V_{ijk}$, whose inclusion is needed
to reproduce the observed binding energies of the three-nucleon system and
the empirical nuclear matter equilibrium properties, consists of the 
Fujita-Miyazawa two-pion exchange potential supplemented
by a purely phenomenological repulsive contribution \ci{uix}.

The predictive power of the dynamical model based on the  hamiltonian
of Eq.(\ref{hamiltonian}) has been extensively tested by computing the energies
of the ground and low-lying excited states of nuclei with $A \leq 12$.
The results of these studies, in which the many body Schr\"odinger
equation is solved {\em exactly} using stochastic methods,
turn out to be in excellent agreement with experimental data \cite{WP}.
Accurate calculations can also be carried out for uniform nuclear matter,
exploiting translational invariance and using the stochastic method
\cite{AFDMC}, the variational approach \cite{Akmal98}, or $G$-matrix perturbation
theory \cite{gmat1,gmat2}.

One of the most prominent features of the NN potential is the 
strongly repulsive core, whose cleanest manifestation is the observed 
saturation of nuclear charge densities.
Due to the presence of the core, the NN potential cannot 
be used to carry out {\em ab initio} microscopic calculations of nuclear observables
using standard perturbation theory. The matrix elements of the interaction hamiltonian 
between eigenstates of the noninteracting system, Fermi gas states in the case of uniform 
nuclear matter, turn out to be very large, or even divergent.

In the $G$-matrix approach the above problem is circumvented replacing 
the bare NN potential with the {\em well behaved} operator $G$, defined  through the Bethe-Goldstone equation
\be
\nonumber
& & \langle i j | G(E) | k l \rangle = G_{ij,kl}(E) \\ 
& = &  v_{ij,kl}  +  \sum_{mn}v_{ij,mn} \frac{Q_{mn}}{E-\epsilon_m-\epsilon_n+i\eta} 
G_{mn,kl}(E) ,
\label{def:Gmatrix}
\ee
where $i \equiv ({\bf k}_i, s_i, t_i)$, ${\bf k}_i$, $s_i$ and $t_i$ being the momentum and the  
spin and isospin projections specifying the i-th single particle state. The Pauli operator $Q_{mn}$ restrict
the sum over intermediate states to those compatible with the exclusion principle, while the 
so-called starting energy $E$ corresponds to the sum of the non-relativistic energies of the 
interacting nucleons. 

The single-particle energy  of a nucleon in the state $i$  is given by 
\beq
\epsilon_i = \frac { k_i^2}{2 m} + Re[U_i] \ ,
\eeq
where $U_i$ describes the mean field felt by the nucleon due to
 its interactions with the other particles of the medium. In the  so-called Brueckner-Hartree-Fock
 approximation, $U_i$  is calculated in the ``on-shell approximation'' through a self-consistent process.
The resulting expression is
\beq
U_i = \sum_{j \in \left\{ F \right\} } \langle ij \mid G(E=\epsilon_i+\epsilon_j) \mid ij \rangle _a \ ,
\eeq
where the sum runs over all occupied states in the Fermi sea  $\left\{ F \right\}$ and the two-nucleon matrix elements 
are properly antisymmetrized.  We note here that the so-called continuous prescription \cite{gmat2} has been adopted for 
the single-particle potential when solving the Bethe-Goldstone equation. As shown in Ref.\cite{baldo1}, the contribution to the energy per particle from three-hole  line diagrams is minimized by this prescription. 

Once a self-consistent solution of the $G$-matrix is achieved, the energy per particle at the two-hole line level takes the form  
\beq
\frac{E}{A} = \frac{3}{5}\ \frac{ k_F^2}{2m} + \frac{1}{2} \sum_{i,j \in \left\{ F \right\} } \langle i j | G(E=\epsilon_i+\epsilon_j) | i j \rangle_a \ ,
\label{E:gmat}
\eeq
where $k_F$ is the Fermi momentum, related to the density through the relation  
 $\rho = \nu k_F^3/6 \pi^2$,  $\nu$ being the spin-isospin degeneracy of the momentum eigenstates 
($\nu =$ 2 and 4 for pure neutron matter and symmetric nuclear matter, respectively). 

In the approach based on correlated wave functions 
one uses the bare potential, $v$, whose non perturbative 
effects are incorporated in the basis states, obtained from the Fermi gas states $|n_{FG}\rangle$ through 
the transformation 
\begin{equation}
|n\rangle = F |n_{FG}\rangle \ .
\end{equation}
The operator $F$, embodying the correlation structure induced by the NN
interaction, is written in the form
\begin{equation}
F=\mathcal{S}\prod_{ij} f_{ij} \  ,
\end{equation}
where $\mathcal{S}$ is the symmetrization operator accounting for the fact that, in general, 
$[f_{ij},f_{ik}] \neq 0$. The two-body correlation functions $f_{ij}$,
whose operator structure reflects the complexity of the NN potential, is written in the 
form
\beq
f_{ij}=\sum_{n=1}^6 f^{n}(r_{ij}) O^{n}_{ij} \ ,
\label{def:corrf}
\eeq
with the $O^n_{ij}$ given by Eq.(\ref{av18:2}).

Within the correlated basis functions (CBF) approach, at two-body cluster level one finds \ci{shannon,bv}
\beq
\frac{E}{A} = \frac{3}{5}\ \frac{k_F^2}{2m} + \sum_{j>i} \langle i j | V_{{\rm eff}} | i j \rangle_a \ ,
\label{E:CBF}
\eeq
where 
\be
\nonumber
\label{veff:2B}
V_{{\rm eff}} & = & \sum_{i < j} f_{ij}^\dagger \left[ -\frac{1}{m} (\nabla^2 f_{ij}) \right. \\
&  &  \ \ \ \ \ \ \ \ \ \ \ \ \ \
\left. - \frac{2}{m} (\bm{\nabla} f_{ij}) \cdot \bm{\nabla} + v_{ij}f_{ij} \right]  \ ,
\ee
and the derivatives act on the relative coordinates.

One would be tempted to exploit the analogies between Eqs.(\ref{E:gmat}) and (\ref{E:CBF}),
to establish a direct link between $G$ and $V_{{\rm eff}}$. Hovewer, determining such a 
connection at operator level is not trivial. 
To see this, just consider that, while all matrix elements
of $G$ involve the bare interaction $v$ only, the matrix elements of $V_{{\rm eff}}$ also include 
 purely kinetic contributions, not containing $v$, which arise from the derivatives of the correlation 
functions. In addition, unlike $V_{{\rm eff}}$, $G$ exhibits an explicit energy dependence. 

The CBF effective interaction, being defined through its ground state expectation 
value, is somewhat limited in scope, with respect to the $G$-matrix effective 
interaction. However, a systematic comparison between the two formalisms 
can be carried out at the level of matrix elements. In this work we will focus
on the matrix elements of the effective interactions in momentum space relevant to 
the calculation of the NN scattering rate in pure neutron matter, whose knowledge 
is required to obtain the transport coefficients within the 
Landau-Abrikosov-Khalatnikov formalism.

We have used the truncated version of the Argonne $v_{18}$
potential referred to as $v_6^\prime$ \cite{V8P}, whose definition only involves 
the static contributions,  i.e. those corresponding to $n \leq 6$,  in Eq.(\ref{av18:2}). 
The CBF effective interaction derived from this potential has been also used 
to obtain weak response of nuclear matter at moderate momentum transfer \cite{shannon,bf}.

For the sake of simplicity, in this work we have neglected the contribution of the 
three nucleon potential appearing in Eq.(\ref{hamiltonian}). 

\section{NN scattering in the nuclear medium}
\label{NNscatt}

\subsection{Kinematics}
\label{kine}

Consider the process in which two nucleons carrying momenta ${\bf k}_1$ and
${\bf k}_2$ scatter to final states of momenta ${\bf k}_1^\prime$ and ${\bf k}_2^\prime$. 
The total energy of the initial state
\beq
E = \frac{{\bf k}_1^2}{2m} + \frac{{\bf k}_2^2}{2m}
\eeq
can be conveniently rewritten in terms of the center of mass and relative momenta, 
${\bf K}~= {\bf k}_1~+~{\bf k}_2$ and ${\bf k}~=~({\bf k}_1~-~{\bf k}_2)/2$, as
\beq
E = \frac{{\bf K}^2}{2M} + \frac{{\bf k}^2}{2\mu} = {\cal E}+ {\cal E}_{{\rm rel}} \ ,
\eeq
with $M = 2m$ and $\mu = m/2$.

In the reference frame in which the center of mass of the system is at rest (CM frame) 
$E=E_{{\rm CM}}={\cal E}_{{\rm rel}}$, while in the lab (L) frame, in which ${\bf k}_2=0$, 
$E=E_{{\rm L}}= 2{\cal E}_{{\rm rel}}$.

The analysis of the NN scattering rates relevant to the calculation of the transport 
coefficients is carried out in the frame in which the Fermi sphere is at rest, often 
referred to as Abrikosov-Khalatnikov (AK) frame. Moreover, in the low-temperature regime, 
in which the results of Ref.\ci{ak} are applicable, scattering processes can only involve 
nucleons with momenta close to the Fermi momentum. Therefore, one can set
\beq
|{\bf k}_1| = |{\bf k}_2| = |{\bf k}_1^\prime| = |{\bf k}_2^\prime| = k_F\ .
\eeq
At energies below pion production threshold the scattering process is elastic, 
so that the requirement of energy conservation 
\be
\nonumber
({\bf k}_1 + {\bf k}_2)^2  & = & 2 k_F^2 (1+\cos \theta) \\ 
 & = & ({\bf k}_1^\prime + {\bf k}_2^\prime)^2 = 2 k_F^2 (1+\cos \theta^\prime) \ ,
\ee
implies that the angle between the momenta of the two nucleons is the same before 
and after the collision. In general, however, the angle $\phi$ between the initial and
final relative momenta, ${\bf k}$ and ${\bf k}^\prime = ({\bf k}_1^\prime - {\bf k}_2^\prime)/2$, defined 
through
\beq
\cos \phi = \frac{ ({\bf k} \cdot {\bf k}^\prime) }{ |{\bf k}||{\bf k}^\prime| }
\label{def:phi}
\eeq
does not vanish. Hence, for any given Fermi momentum, i.e. for any given matter density, the 
scattering process in the AK frame is specified by the center of mass energy 
\beq
{\cal E}_{{\rm AK}} = \frac{k_F^2}{2m} (1 + \cos \theta ) \ , 
\eeq
and the two angles $\theta$ and $\phi$. 

As the NN scattering cross section is often evaluated in the CM frame, it is 
convenient to establish a relationship between kinematical variables in the CM and AK frames.
Exploiting the frame invariance of the relative energy we easily obtain
\beq
E_{{\rm CM}} = \frac{k_F^2}{2m} (1 - \cos \theta ) \ , 
\label{def:ECM}
\eeq
while the center of mass scattering angle $\theta_{{\rm CM}}$
can be indentified with $\phi$, defined in Eq.(\ref{def:phi}).

\subsection{Cross section}
\label{xsec}

In both the $G$-matrix and CBF efffective interaction approaches, the NN cross 
section in matter at density $\rho$, can be written in the form 
\beq
\frac{d \sigma}{d \Omega_{{\bf k}^\prime}} = \frac{ {m^\star}^2 }{16 \pi^2} \ 
\sum_{S M M'} |{\cal M}_{S}^{M M'} (\theta,\phi)|^2  \ ,
\label{xsec:1}
\eeq
where $m^\star$ is the nucleon effective mass,  and the transition amplitude in the channel
of total spin $S$ and initial and final spin projections $M$ and $M'$, $\cal{M}_S^{M M'}(\theta,\phi)$, 
involves the matrix elements of either $G$ or $V_{{\rm eff}}$
between Fermi gas states. 

Numerical calculations of the cross sections are carried out 
expanding $\cal{M}_S^{M M'}(\theta,\phi)$ in partial waves. 
In the case of pure neutron matter, in which the total isospin of the interacting 
pair is $T=1$, the expansion only involves partial waves of even (odd) angular momentum 
in spin singlet (triplet) states. 

The
 matrix elements of the CBF effective interaction  can be written in the form
\beq
M_{\ell \ell^\prime}^{SJ}(k) = \frac {2}{\pi} \int r^2 dr j_\ell(kr) 
\langle \ell^\prime S J | V_{{\rm eff}} | \ell S J \rangle j_{\ell^\prime}(kr) \ .
\label{matel}
\eeq
In the above equation, $\ell$  and $J$ denote orbital  and 
total angular momentum, respectively,  $j_\ell$ is the spherical Bessel function,  
$| \ell S J \rangle$ is the spin-angle state and $r$ is the magnitude of the 
relative distance. 
Note that, due to  the presence of the tensor operator of Eq.(\ref{def:S12}), the 
NN potential couples states of different orbital angular momentum. The above matrix
elements are directly comparable with those obtained from the partial wave expansion of 
the $G$-matrix (see Eq.(\ref{def:Gmatrix})). 
\begin{figure}[hbt]
\centerline
{\epsfig{figure=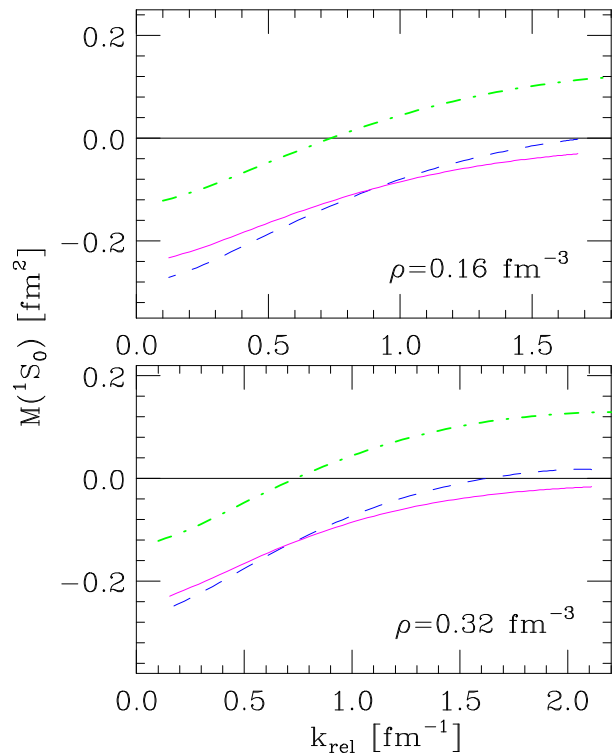,angle=000,width=8.0cm}}
\caption{ (Color online)
Top panel: $M(^1S_0)$ matrix element of the CBF (solid lines) and $G$-matrix
(dashed lines) effective interactions ($M_{00}^{00}$ of Eq.(\ref{matel})) for neutron matter at
nuclear matter equilibrium density, as a function of relative momentum.
For comparison, the dot-dash line shows the result corresponding to the bare
$v_6^\prime$ potential. Bottom panel: same as the top panel, but for density
$\rho = 2 \rho_0$.
\label{matel0} }
\end{figure}

In Fig. \ref{matel0} we show the matrix element $M(^1S_0)~=~M_{00}^{00}$, evaluated 
at nuclear matter equilibrium density, $\rho_0 = 0.16$ fm$^{-3}$ (top panel),  and 
$2\rho_0$ (bottom panel), in the kinematical setup described in Section \ref{kine}.
The solid and dashed lines correspond to the matrix elements of the CBF and $G$-matrix
effective interactions, respectively, while the dot-dash line has been obtained replacing $V_{{\rm eff}}$ 
with the bare $v_6^\prime$ potential. Renormalization of the NN interaction, carried out either solving the 
Bethe-Goldstone equation or modifying the basis states, appears to have a strong impact on the matrix elements, which 
become more attractive with respect to the matrix elements of the bare interaction.  
On the other hand,  
CBF and $G$-matrix approaches yield rather similar results in the considered range of densities and
momenta.

\begin{figure}[htb]
\centerline
{\epsfig{figure=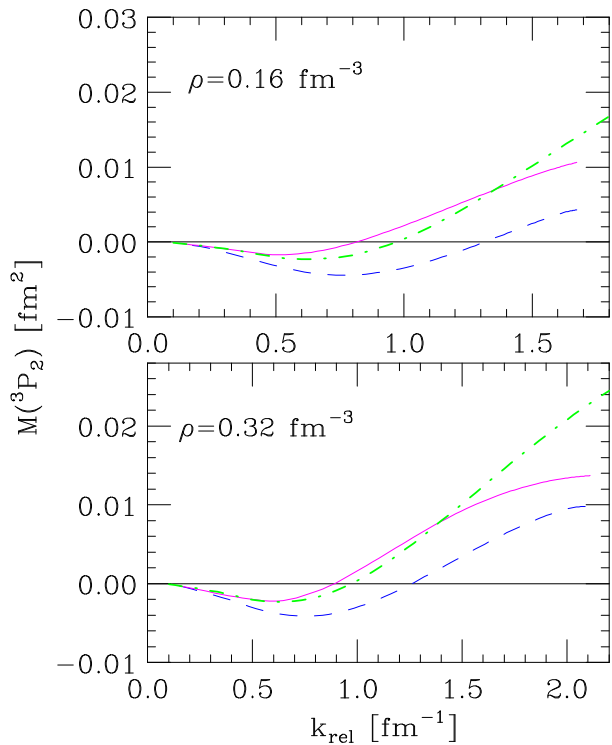,angle=000,width=8.0cm}}
\caption{ (Color online)
Same as in Fig. \ref{matel0}, but for the $M(^3P_2)$ matrix elements.
\label{matel1} }
\end{figure}

The matrix elements $M(^3P_2)=M^{12}_{11}$ are shown in Fig. \ref{matel1}. It appears that
in this channel the results obtained using the CBF are closer to those corresponding to 
the bare interaction, while being appreciably different from the $G$-matrix results.
Note, however, that the matrix elements corresponding to the $^3P_2$ channel are 
over one order of magnitude smaller than those corresponding to the $^1S_0$ channel.

It has to be pointed out that the density dependence of  the matrix elements is rather mild. 
In the CBF approach the density dependence arises from the two-body correlation functions, while
in the G-matrix it comes through the presence of the Pauli operator 
and the single particle potentials appearing in the denominator of Eq.(\ref{def:Gmatrix}). 
In addition, one should also take into account the density dependence associated with the 
starting energy $E$, as the matrix elements 
reported in the Figs. \ref{matel0} and \ref{matel1}  have been computed
at twice the Fermi energy of the corresponding density. 

\begin{figure}[htb]
\centerline
{\epsfig{figure=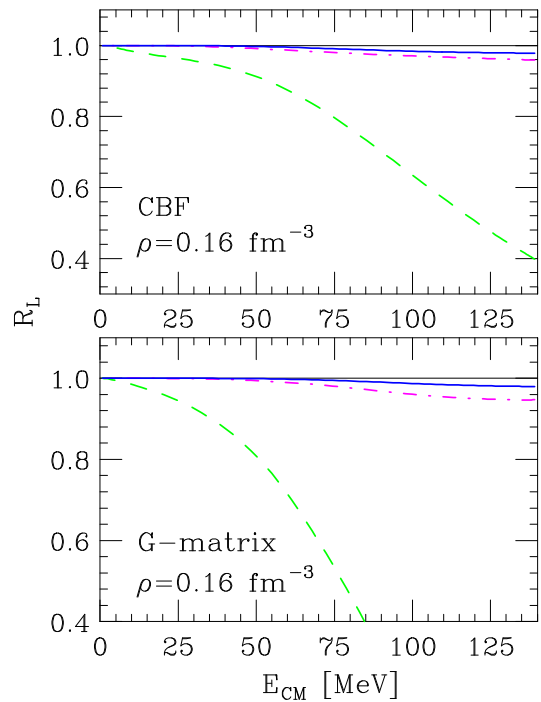,angle=000,width=7.0cm}}
\caption{ (Color online)
Energy dependence of the ratio $R_L$, defined by Eq.(\ref{ratio}).
The dashed, dot-dash and solid lines have been obtained including
the contributions of states of angluar momentum $\ell$ up to 0, 1 and 2.
The top and bottom panels correspond to CBF and $G$-matrix effective interactions,
respectively.
\label{convergence} }
\end{figure}

The convergence of the partial wave expansion is illustrated in Fig, \ref{convergence}, 
showing the ratio
\beq
\label{ratio}
R_{L} = \frac{ 1 }{ \sigma_{{\rm tot}} } \  \sum_{\ell=0}^L \ \sigma_{{\rm tot}}^\ell  \ ,
\eeq
as a function of CM energy (see Eq.(\ref{def:ECM})). In the above equation 
 $\sigma_{{\rm tot}}$ is the total in-medium neutron-neutron cross section, while
$\sigma_{{\rm tot}}^\ell$ denotes the contribution of the $\ell$-th partial wave. 
All cross sections have been evaluated in the kinematical set up relevant to the calculation of the 
transport coefficients, discussed in Section \ref{kine}.
 The definition obviously implies that, as $L \to \infty$, $R_L  \to 1$.

The results of Fig. \ref{convergence} show that the total cross sections obtained 
including only the partial waves with $\ell =0$ and 1, corresponding to the spin-singlet and
spin-triplet states of lowest angular momentum, is within less than 5\% of the fully 
converged result. Comparison between the top and bottom panels, corresponding to $CBF$ and
$G$-matrix effective interactions, also shows that the two approaches lead to a similar qualitative
behavior, although the $G$-matrix $R_0$ exhibit a somewhat steeper energy dependence.

\begin{figure}[htb]
\centerline
{\epsfig{figure=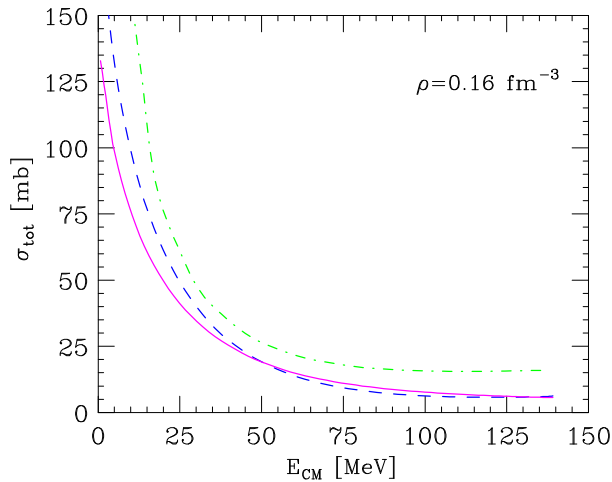,angle=000,width=8.0cm}}
\caption{ (Color online)
Total in-medium neutron-neutron cross section in neutron matter, computed at 
nuclear matter equilibrium density
as a function of energy in the center of mass frame.
The solid and dashed lines have been obtained using the CBF and $G$-matrix effective
interactions, respectively, in the kinematical setup discussed in Section III A.
 For comparison, the dot-dash line shows the free-space cross section,  
obtained from the $t$-matrix associated with the $v_6^\prime$ potential.
\label{sigmatot} }
\end{figure}

In Fig. \ref{sigmatot} we compare the total neutron-neutron cross section at nuclear matter 
equilibrium density, computed in the kinematics of Section \ref{kine}, as a 
function of $E_{{\rm CM}}$. The difference between the cross sections obtained using the CBF 
(solid line) and $G$-matrix (dashed line) approaches does not exceed $\sim$ 20 \% 
at $E_{{\rm CM}} > $ 10 MeV. On the other hand, the 
screening effect due to the presence of the nuclear medium, illustrated by the difference
between the solid and dashed lines and the dot-dashed one, corresponding to the free-space cross section 
obtained from the $t$-matrix associated with the $v_6^\prime$ potential, turns out to be large. 
At $E_{{\rm CM}} > $ 100 MeV, where the CBF and $G$-matrix results are very close to one another, the in 
medium cross section turns out to be quenched by a factor $\sim$ 3.

\section{Transport coefficients}
\label{transport}

\subsection{Abrikosov-Khalatnikov formalism}
\label{formalism}

The theoretical description of transport properties of normal Fermi liquids
is based on Landau theory \cite{baym-pethick}.
Working within this framework and including the leading
term in the low-temperature expansion, Abrikosov and Khalatnikov \cite{ak} obtained
approximate expressions for the shear viscosity and the thermal conductivity.
Let us consider viscosity, as an example. The AK result reads
\beq
\label{eta_AK}
\eta_{AK} = \frac{1}{5}\rho m^\star v^2_F \tau \,\frac{2}{\pi^2(1-\lambda_\eta)} \ ,
\eeq
where $v_F=k_F/m^\star$ is the Fermi velocity and $m^\star$ and $\tau$ denote
the quasiparticle effective mass and lifetime, respectively. The latter can be
written in terms of the angle-averaged scattering probability, $\langle \cal{W} \rangle$, with 
(see Eq.(\ref{xsec:1})) 
\beq
{\cal W}(\theta,\phi) = \sum_{S M M^\prime} |{\cal M}_S^{{M M^\prime}} (\theta,\phi)|^2 \ ,
\eeq
according to
\beq
\label{tau_AK}
\tau T^2 = \frac{8\pi^4}{{m^*}^3}\ \frac{1}{\langle {\cal W} \rangle} \ ,
\eeq
where $T$ is the temperature and
\beq
\label{Wavg}
\langle {\cal W} \rangle = \int \frac{d\Omega}{2\pi}\ \frac{{\cal W}(\theta,\phi)}
{\cos{(\theta/2)}} \ .
\eeq
Note that, as pointed out in Section \ref{kine}, the scattering process involves quasiparticles on the Fermi surface.
As a consequence, for any given density $\rho$, ${\cal W}$ depends only on
the angular variables $\theta$ and $\phi$.
Finally, the quantity $\lambda_\eta$ appearing in Eq.(\ref{eta_AK}) is defined as
\beq
\lambda_\eta = \frac{\langle {\cal W} [ 1-3\sin^4{(\theta/2)}\sin^2{\phi} ] \rangle}
{\langle {\cal W} \rangle} \ .
\label{lambda:eta}
\eeq

The exact solution of the equation derived in Ref. \cite{ak}, obtained by
Brooker and Sykes \cite{sb1}, reads
\be
\nonumber
\eta & = & \eta_{AK} \frac{1-\lambda_\eta}{4} \\ 
& \times &
 \sum_{k=0}^\infty \frac{4k+3}{(k+1)(2k+1)[(k+1)(2k+1)-\lambda_\eta]} \ ,
\label{eta_sb}
\ee
the size of the correction with respect to the result of Eq.(\ref{eta_AK}) being
$0.750 < (\eta/\eta_{AK}) < 0.925$.

The expression of the thermal conductivity, $\kappa$, is obtained following the 
same procedure, the only difference being that in this case the leading term in the 
low energy expansion is linear, rather than quadratic, in the inverse temperature $T^{-1}$.
The resulting expression is
\be
\label{cappa_sb}
\nonumber
\kappa & = & \kappa_{AK} \frac{3 - \lambda_\kappa}{4} \\
& \times &  \sum_{k=0}^\infty \frac{4k+5}{(k+1)(2k+3)[(k+1)(2k+3)-\lambda_\kappa]} \ ,
\ee
where
\beq
\kappa_{AK} = \frac{1}{T} \ \frac{8}{3} \ \frac{k_F^3}{{m^\star}^4} \frac{2 \pi^2}{\langle {\cal W} \rangle (3-\lambda_\kappa)}  
\eeq
and
\beq
\label{lambda:cappa}
\lambda_\kappa = \frac{\langle {\cal W} (1+2 \cos \theta) \rangle}
{\langle {\cal W} \rangle} \ . 
\eeq

In this case the correction to the AK result turns out to be larger. From Eq.(\ref{lambda:cappa}) 
it follows that $-1 < \lambda_\kappa < 3$, implying in turn (see Eq.(\ref{cappa_sb})) $0.417~<~(\kappa/\kappa_{AK})~<~0.561$.
\subsection{Results}
\label{results}

\begin{figure}[hbt]
\centerline
{\epsfig{figure=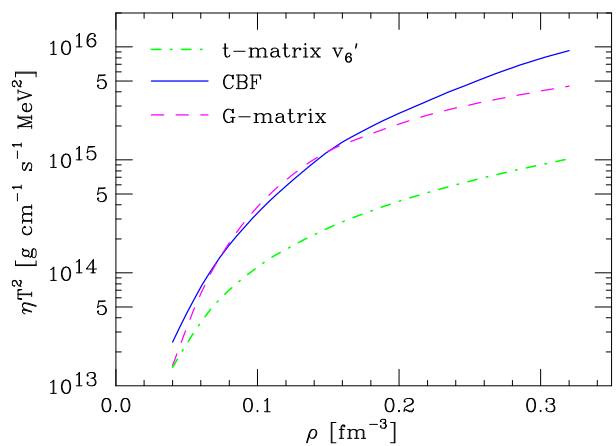,angle=000,width=8.0cm}}
\caption{ (Color online)
Density dependence of the $T$-independent quantity $\eta T^2$ in pure neutron matter. The solid and dashed lines  
correspond to the CBF and $G$-matrix effective interactions, respectively, while the dot-dash line shows the 
results obtained from the $t$-matrix associated with the $v_6^\prime$ potential.
\label{etaT2} }
\end{figure}

\begin{figure}[hbt]
\centerline
{\epsfig{figure=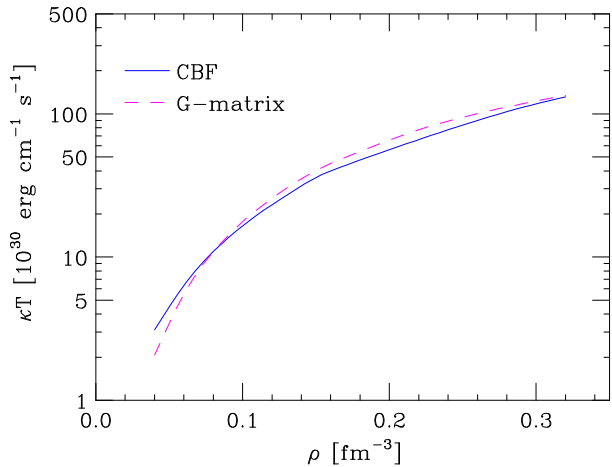,angle=000,width=8.0cm}}
\caption{ (Color online)
Density dependence of the $T$-independent quantity $\kappa T$ in pure neutron matter. The solid and dashed lines have been 
obtained using the CBF and $G$-matrix effective interactions, respectively.
\label{cappaT} }
\end{figure}

Figures \ref{etaT2} and \ref{cappaT} show the $T$-independent 
 quantities
$\eta T^2$ and $\kappa T$, respectively, as a function of density. The calculations have been carried out  using 
the formalism described in the previous Section and the scattering probabilities ${\cal W}(\theta,\phi)$ obtained from both 
the $G$-matrix and CBF effective interactions, which have been computed at zero temperature. 
For comparison, in Fig. \ref{etaT2} we also display, by the dot-dash line, the results obtained from the free-space scattering 
probability, computed  using the $t$-matrix associated with the bare $v_6^\prime$ potential. 

As the ratios $m^\star/m$ resulting from the two approaches, CBF and G-matrix,  turn out to be rather close to one another (the difference
never exceeds few percent in the density range shown in Figs. \ref{etaT2} and \ref{cappaT}) all calculations
have been carried out using the CBF effective masses.

Figure \ref{etaT2} clearly indicates that medium modifications of the NN scattering cross sections play a critical role, leading
to a  dramatic enhancement of the viscosity. A similar effect is observed in the case of thermal conductivity \cite{BFFV}.

Comparison between Fig. \ref{etaT2} and Fig. \ref{cappaT} shows that, while the density dependence of the thermal conductivity
resulting from the two approaches looks remarkably similar, in the case of viscosity sizable discrepancies occur at  densities larger
than nuclear matter equilibrium density.

This feature can be easily explained considering the different angular dependence of the integrands in Eqs.(\ref{lambda:eta}) 
and (\ref{lambda:cappa}), which determine the functions $\lambda_\eta$ and $\lambda_\kappa$, respectively. 
In the case of $\lambda_\kappa$, ${\cal W}(\theta,\phi)$ is weighted with a function that only 
depends on $\theta$. As a consequence, the result only depends on the $\phi$-integrated scattering probability, which is 
trivially related to the {\em total} cross section (see Eq.(\ref{xsec:1})) at energy $E_{CM}$ given by Eq.(\ref{def:ECM}). 
Hence, the similar behavior exhibited by the two curves of Fig. \ref{cappaT} merely reflects the fact that the total cross
sections obtained from the $G$-matrix and CBF approaches turn out to be close to one another (see Fig. \ref{sigmatot}). 
On the other hand, in the right hand side of Eq.(\ref{lambda:eta}) the scattering probability is weighted with a 
factor that emphasizes the differences in the $\phi$ dependence of the CBF and $G$-matrix $\cal{W}(\theta,\phi)$. 

We have verified that the large discrepancy between the solid and dashed curves of Fig. \ref{etaT2} at high density
is in fact ascribable to $\lambda_\eta$. The $\lambda_\eta$-independent quantities $\eta_{AK}(1-\lambda_\eta)/4$ 
obtained from the CBF and G-matrix approaches turn out to be within less than 5 \% of one another at 
$\rho = 0.32$ fm$^{-3}$. On the other hand, removal of the $\lambda_\kappa$ dependence in $\kappa_{AK}$ does 
not produce any significant effects. 
 
\section{Conclusions}
\label{conclusions}

Many body theory provides a fully consistent framework, suited to construct effective interactions starting from 
highly realistic models of nuclear dynamics. In this work, we have employed the effective interactions resulting 
from the $G$-matrix and CBF approaches to compute the in-medium NN cross sections, which are needed to obtain 
the transport coefficient within the Landau-Abrikosov-Khalatnikov formalism. 

The calculations have been carried out using the truncated $v_6^\prime$ form of the NN potential of Ref. \cite{av18}.
The effects of three- and many-body forces, though being known to be sizable at large density, have not been 
taken into account. The results of Ref. \cite{bv} show that many-body forces give rise to  
a change of the shear viscosity of less than 10~\% at $\rho~\lsim$~0.32~fm$^{-3}$. Hence, their 
inclusion is not likely to significantly affect the main conclusions of our work.

The approach based on effective interactions allows one to consistently take into account screening effects 
arising from short range NN correlations, that lead to a large decrease of the scattering cross section.
As a consequence, the shear viscosity and thermal conductivity obtained from the effective interactions turn 
out to be much larger than the corresponding quantities computed using the bare NN potential.

Our work, showing that the results of the $G$-matrix and CBF schemes are in reasonable agreement with 
one another, suggests that as long as the effective interaction is based on a {\em realistic} NN potential, strongly 
constrained by the large data set of NN scattering phase shifts, the model dependence associated with the 
many body approach employed is not critical. 

On the other hand, it has to be pointed out that Skyrme effective interactions (for a recent 
and comprehensive discussion the application of the Skyrme approach to nuclear matter and neutron 
stars see Ref.\cite{skyrme1}), mainly constructed by fitting bulk properties of nuclear matter, predict in-medium 
NN cross sections whose behavior is significantly different 
from the one predicted by the $G$-matrix and CBF effective interactions. As a result, the values of the viscosity and 
thermal conductivity  coefficients computed using  Skyrme effective interactions turn out to be much lower 
than those shown in Figs. \ref{etaT2} and \ref{cappaT}. For example, using the SLya 
effective interaction, adjusted to reproduce the the microscopically derived EOS of neutron and nuclear matter \cite{skyrme2},
one finds at nuclear matter equilibrium density $\eta T^2 \sim 6\times10^{13}$ g cm$^{-1}$ s$^{-1}$ MeV$^2$ 
and $\kappa T \sim 4\times10^{30}$ erg cm$^{-1}$ s$^{-1}$, to be 
compared to $\sim 1.4\times10^{15}$ g cm$^{-1}$ s$^{-1}$ MeV$^2$ and 
$\kappa T \sim 4\times10^{31}$ erg cm$^{-1}$ s$^{-1}$
obtained from the $G$-matrix and CBF formalisms.
While the Skyrme approach has proved to be very useful in many contexts, these results 
suggest that the determination of the transport properties of nuclear matter requires
effective interactions providing a quantitative account of the observed NN scattering data in 
the limit of vanishing density.
 
\acknowledgments
This work was partially supported  by FEDER/FCT (project CERN/FP/83505/2008),  
Consolider-Ingenio 2010 (programma CPAN CSD2007-00042 and grant FIS2008-01661), 
MEC/FEDER (project 2009SGR-1289) and the European Science Foundation research networking 
program COMPSTAR. MV gratefully acknowledges the hospitality of the Departament d'Estructura i 
Constituents de la Mat\`eria, Barcelona. 



\end{document}